\documentclass[
    ,final            
  ]
  {aipproc}

\layoutstyle{6x9}

\newcommand{\gthree}{${\rm G}_{300}^{\rm B180 }$}
\newcommand{\hv}{HV}
\newcommand{\msun}{${\rm M}_{\odot}$}

\newcommand{\mevt}{{\rm MeV/fm}^3}
\newcommand{\nuk}{{\nu_{\rm K}}}

\begin{document}

\title{Neutron star interiors and the equation of state of ultra-dense matter}



\author{F. Weber}{
  address={Dept. of Physics, San Diego State University,
  5500 Campanile Drive, San Diego, CA 92182, USA}
}

\author{R. Negreiros}{
  address={Dept. of Physics, San Diego State University,
  5500 Campanile Drive, San Diego, CA 92182, USA}
}

\author{P. Rosenfield}{
  address={Dept. of Physics, San Diego State University,
  5500 Campanile Drive, San Diego, CA 92182, USA}
}

\author{Andreu Torres i Cuadrat}{
        address={Physics Department, Universitat Autonoma de Barcelona, Spain}
}

\begin{abstract}
There has been much recent progress in our understanding of quark
matter, culminating in the discovery that if such matter exists in the
cores of neutron stars it ought to be in a color superconducting
state. This paper explores the impact of superconducting quark matter
on the properties (e.g., masses, radii, surface gravity, photon
emission) of compact stars.
\end{abstract}

\maketitle


Exploring the composition of matter inside compact stars has
become a forefront area of modern physics
\cite{glen97:book,weber99:book}. Despite the progress that was made over
the years, the physical properties of the matter in the ultra-dense
core of compact stars is only poorly known. Recently it has been
theorized that, if quark matter exists in the core, it ought to be a
color superconductor \cite{rajagopal01:a,alford01:a}.  This paper
reviews the consequences of color superconducting quark matter cores
on the properties of neutron stars. The study is based on three
\begin{figure}[h]
\includegraphics[width=0.6\textwidth,angle=0,clip]{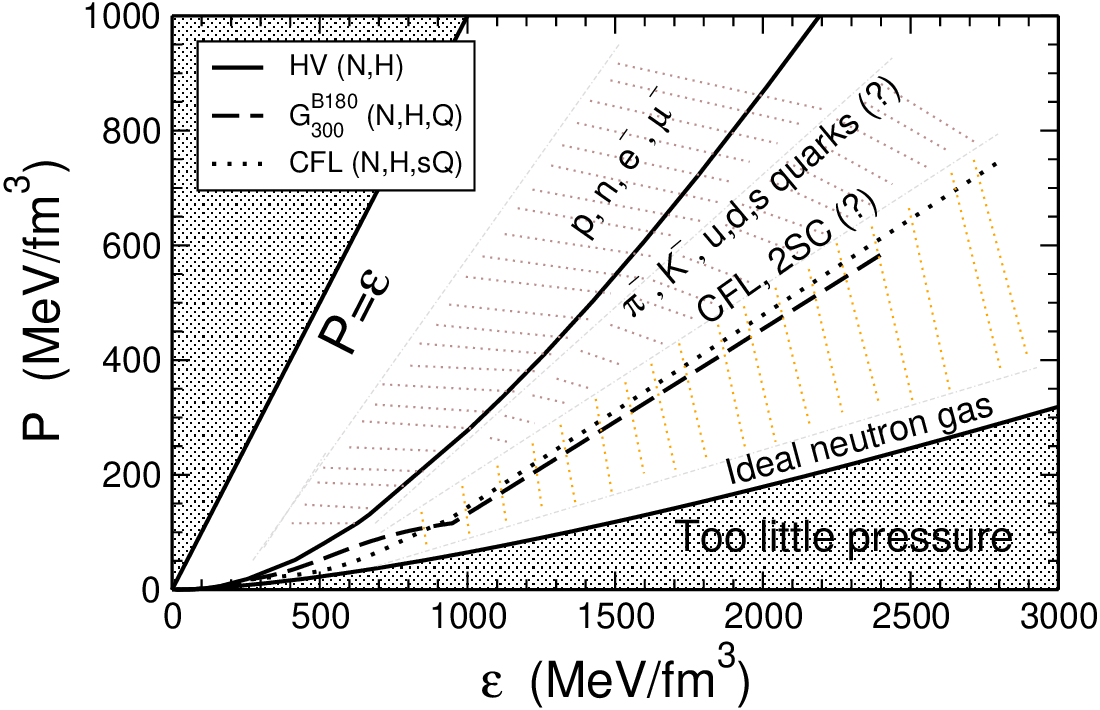}
\caption{Eos considered in this study. The shaded areas 
reflect the uncertainties in the eos originating from different many-body
treatments and competing assumption about the particle composition.}
\label{eos}
\end{figure}
sample models for the nuclear equation of state (eos).  The first model,
\hv \cite{glen85:b}, treats the core matter as made of conventional
hadronic particles (nucleons and hyperons) in chemical equilibrium
with leptons (electrons and muons). The second eos, \gthree
~\cite{glen97:book}, additionally accounts for non-superconducting
quark matter. Finally, the third model accounts for quark matter
in the superconducting color-flavor locked (CFL) phase
\cite{alford03:b}. Figure \ref{eos} shows these eos graphically.
\begin{figure}[ht]
\includegraphics[width=0.6\textwidth,angle=0,clip]{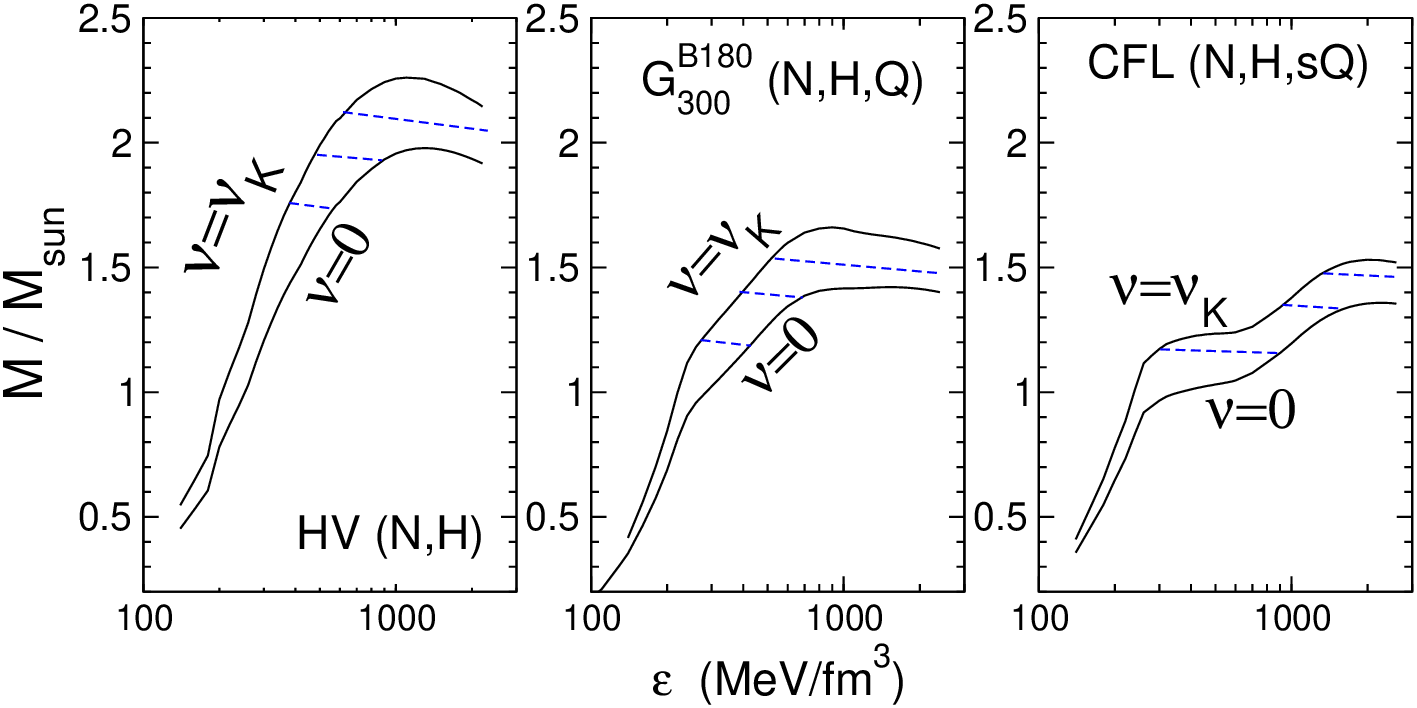}
\caption{Mass--central energy density relations for the three
         model stars of this study.}
\label{mvse}
\end{figure}

\begin{figure}[ht]
\includegraphics[width=0.4\textwidth,angle=0,clip]{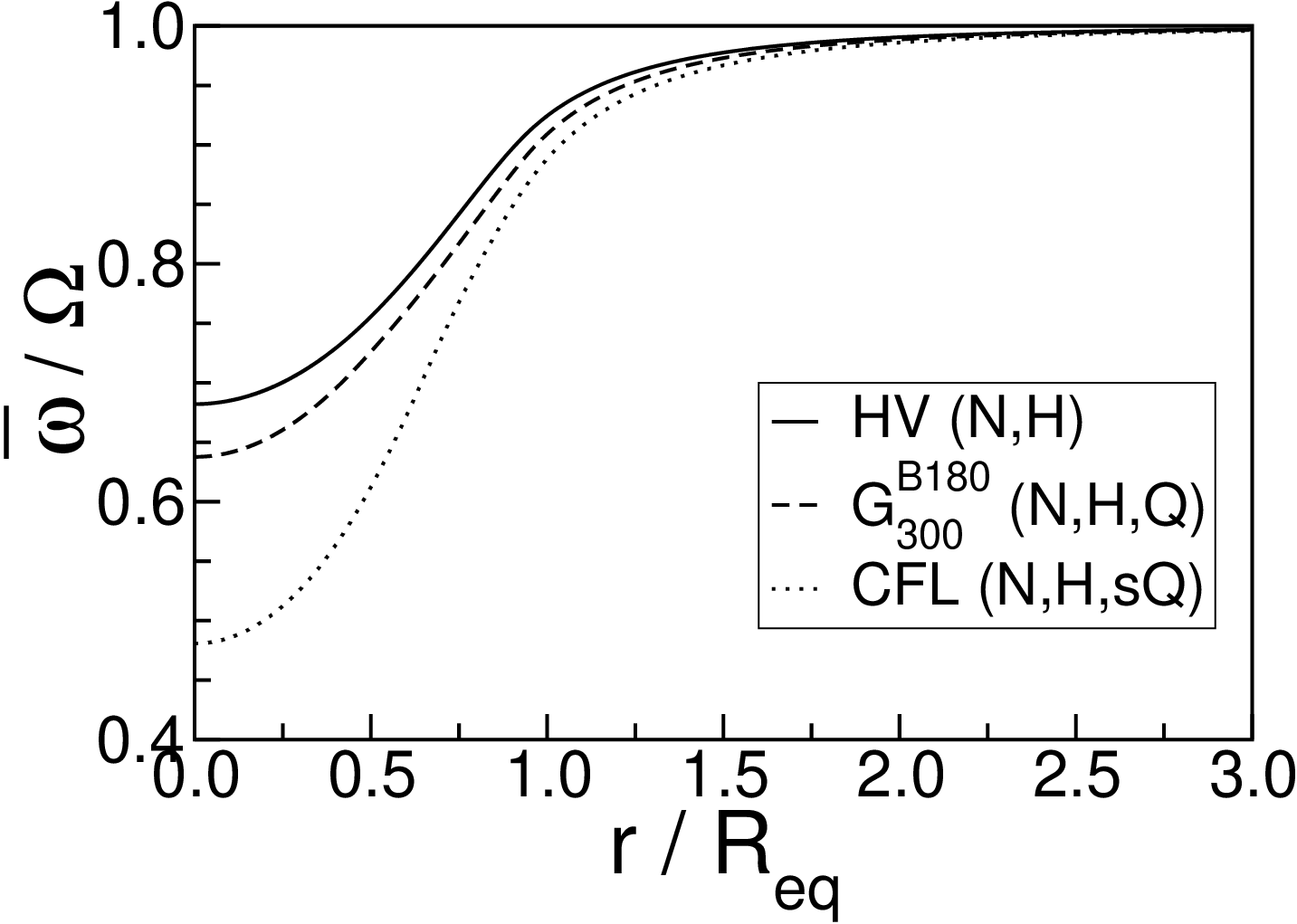}
\caption{Lense--Thirring effect caused by $\sim 1.4$ \msun stars rotating at 2~ms.}
\label{fd} 
\end{figure}

Figure \ref{mvse} shows the evolutionary (constant stellar baryon
number, $A$) paths that isolated rotating neutron stars would follow
during their stellar spin-down caused by
the emission of magnetic dipole radiation and a wind on $e^+$--$e^-$
pairs. Figure \ref{mvse} reveals that CFL stars may spend considerably
more time in the spin-down phase than their competitors of the same
mass.  Figure \ref{fd} shows the general relativistic effect of frame
dragging \cite{glen97:book,weber99:book,weber05:a}, which is
considerably more pronounced for the CFL stars because of their much 
greater densities. This may be of great
importance for binary millisecond neutron stars in their final
accretion stages, when the accretion disk is closest to the neutron
star.
Table \ref{tab:redshifts} summarizes the impact of strangeness on
several intriguing properties of non-rotating as well as rotating
neutron stars. The latter spin at their respective Kepler frequencies.
One sees that the central energy density, $\epsilon_{\rm c}$, spans a
\begin{table}[htb]
\begin{tabular}{ccccccc} 
\hline
 & ~\hv~ & ~\gthree~ & ~CFL~ & ~\hv~ & ~\gthree~ & ~CFL~ \\
& $\nu$ = 0  &$\nu = 0$  &$\nuk = 0$ & $\nuk$ = 850~Hz
& $\nuk$ = 940~Hz  &$\nuk$ = 1400~Hz \\
\hline 
$\epsilon_{\rm c}~(\mevt)$ & 361.0  & 814.3  &2300.0 &280.0  &400.0 &1100.0  \\
$I~{\rm (km^{3})}$         & 0      & 0      & 0      & 223.6   & 217.1   & 131.8   \\
$M$~(\msun)                  & 1.39   & 1.40   & 1.36   & 1.39    & 1.40    & 1.41    \\
$R$~(km)                     & 14.1   & 12.2   & 9.0    & 17.1    & 16.0    & 12.6    \\
$Z_{\rm p}$           & 0.1889 & 0.2322 & 0.3356 & 0.2374  & 0.2646  & 0.3618  \\
$Z_{\rm F}$     & 0.1889 & 0.2322 & 0.3356 & $-$0.1788 & $-$0.1817 & $-$0.2184 \\
$Z_{\rm B}$             & 0.1889 & 0.2322 & 0.3356 & 0.6046  & 0.6502  & 0.9190  \\
$g_{\rm {s,14}~ (cm/s^{2})}$ & 1.1086 & 1.5447 & 3.0146 & 0.7278  & 0.8487  & 1.4493  \\
$T/W$                      & 0      & 0      & 0      & 0.0894  & 0.0941  & 0.0787  \\
$BE$~(\msun)               & 0.0937 & 0.1470 & 0.1534 & 0.0524  & 0.1097  & 0.1203  \\
$V_{\rm eq}/c$           & 0      & 0      & 0      & 0.336   & 0.353   & 0.424   \\
\hline
\end{tabular}
\caption{Properties of neutron stars composed of nucleons and hyperons (\hv), 
nucleons, hyperons, normal quarks (\gthree), and nucleons,
hyperons,  superconducting quarks (CFL)
\cite{weber99:book}.}
\label{tab:redshifts} 
\end{table}
very wide range, depending on particle composition. The surface
redshift is of importance since it is connected to observed neutron
star temperatures through the relation $T^\infty/T_{\rm eff} = 1/(1+Z)$.
CFL quark stars may have redshifts that are up to 50\% higher than
\begin{figure}[htb]
\includegraphics[width=0.85\textwidth,angle=0,clip]{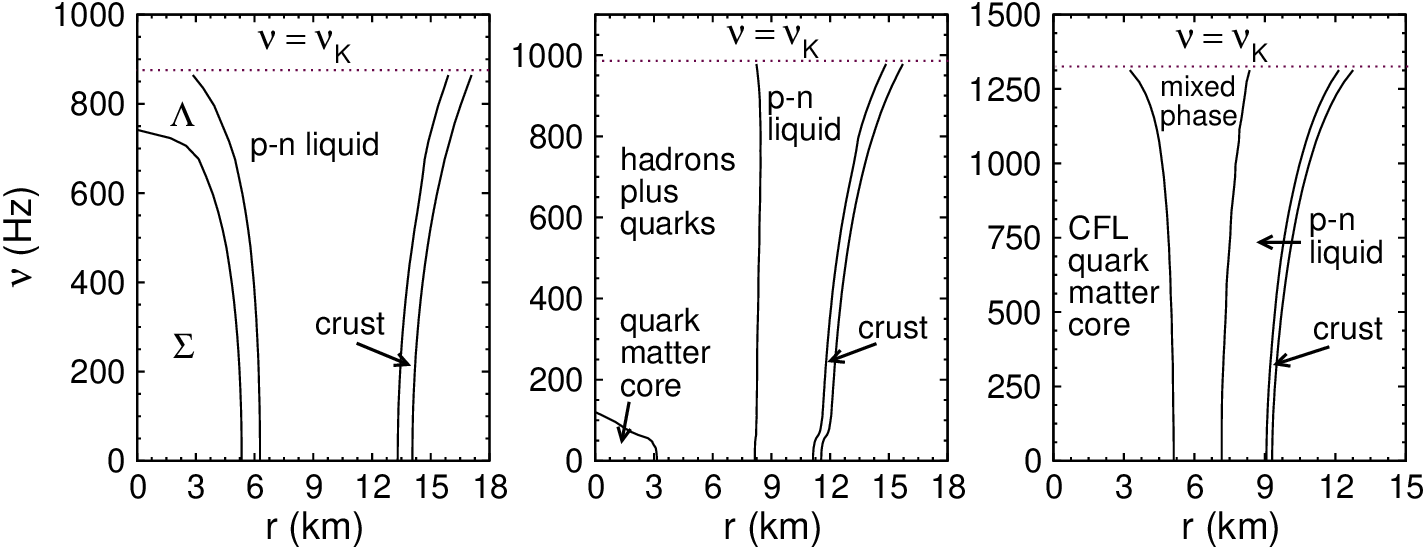}
\caption{Particle profiles of the neutron stars of our
                collection.}
\label{prof}
\end{figure}
those of conventional stars. Finally, we also show in Table
\ref{tab:redshifts} the surface gravity of stars, $g_{{\rm s},14}$
\cite{bejger04:a}, which again may be up to 50\% higher for CFL
stars. The other quantities listed are the rotational kinetic energy
in units of the total energy of the star, $T/W$,
the stellar binding energy, $BE$, and the rotational velocity of a
particle at the star's equator \cite{weber99:book}.


\begin{theacknowledgments}
This work is supported by the National Science Foundation under Grant
PHY-0457329, and by the Research Corporation.
\end{theacknowledgments}



\end{document}